\documentclass[10pt,letterpaper]{article}
\usepackage{opex3}
\usepackage{color}
\usepackage{amsmath}
\usepackage{amsfonts}
\usepackage{amssymb}
\usepackage{cite}
\usepackage{color}
\usepackage{verbatim}
\usepackage{epstopdf}

\begin{document}

\title{Transient and steady-state entanglement mediated by three-dimensional plasmonic waveguides}

\author{S. Ali Hassani Gangaraj, Andrei Nemilentsau, George W. Hanson}

\address{Department of Electrical Engineering and Computer Science, University of Wisconsin-Milwaukee, Milwaukee, Wisconsin 53211, USA}

\email{Hassani4@uwm.edu} %% email address is required

\author{Stephen Hughes}
\address{Department of Physics, Engineering Physics and Astronomy, Queen's University, Kingston, Ontario, Canada K7L 3N6}

\begin{abstract}
Entanglement between two qubits (two level atoms) mediated by surface plasmons in three-dimensional plasmonic waveguides is studied using a quantum master equation formalism. Two types of waveguides, a nanowire and a V-shaped channel cut in a flat metal plane, are considered. The Green functions for the waveguides, which rigorously describes the dissipative qubit environment, are calculated numerically using a direct finite-difference time-domain (FDTD) solution of Maxwell's equations. Finite-length effects are shown to play a crucial role in enhancing entanglement, and resonant-length plasmonic waveguides can provide higher entanglement between qubits than infinite-length waveguides. It is also shown that coupling slots can improve entanglement via stronger qubit-waveguide coupling, for both the infinite- and finite-waveguide cases.   
\end{abstract}

\ocis{(000.0000) General.} % REPLACE WITH CORRECT OCIS CODES FOR YOUR ARTICLE, MINIMUM OF TWO; Avoid using the OCIS codes for “General” or “General science” whenever possible.
%For a complete list of OCIS codes, visit: http://www.opticsinfobase.org/submit/ocis/

%%%%%%%%%%%%%%%%%%%%%%% References %%%%%%%%%%%%%%%%%%%%%%%%%

%%%%%%%%%%%%%%%%%%%%%%%%%%  body  %%%%%%%%%%%%%%%%%%%%%%%%%%
\section{Introduction}

Generating, preserving and controlling entanglement between two quantum systems is of great importance for quantum communications, teleportation, metrology, cryptology, computing and other operations involving quantum bits (qubits) \cite{Aolita}. A fundamental problem facing practical applications of entanglement is environment-induced dissipation of the quantum system. The traditional view \cite{Ryszard} holds that dissipation is necessarily detrimental as it leads to decoherence, and thus to entanglement decay. However, recently it has been realized that dissipation may have a positive effect as well. In particular, it has been demonstrated that by engineering dissipation of the environment, quantum systems can be driven into the desired entangled steady state (SS), encoding the outcome of quantum computations \cite{Verstraete,Barreiro,Lin}. In this case, the depopulation of the quantum system is compensated by constant pumping with an external electromagnetic field, and properties of the final states depend on the environment to which the systems are coupled. Therefore, engineering of the environment seen by qubits is of crucial importance for applications in quantum optics. 

Of particular interest are structured photonic reservoirs, i.e. structures that have strongly inhomogeneous spatial or spectral distributions of their photonic density of states, as they offer an unprecedented level of control over emitter decay dynamics \cite{Walther,Anger,Akimov,Nemilentsau,Forati}. The efficient dissipative generation of entanglement has been theoretically predicted and experimentally demonstrated for qubits in optical cavities \cite{Kastoryano,Casabone,Reimann} and photonic crystal systems \cite{Yao09,Wolters14,Angelatos2014}. In particular, it was reported \cite{Kastoryano} that dissipative preparation of entanglement between qubits in high-finesse optical cavities offers significant advantages for fidelity improvement over the protocols based on unitary dynamics; decreasing the error by an order of magnitude requires only an order of magnitude increase of the cavity finesse in the former case, compared to a two order of magnitude increase of the finesse required in the latter case. Preparation of entangled states with fidelity (with respect to the Bell state $\Psi^+$) exceeding 91.9$\%$ was experimentally demonstrated \cite{Casabone}. Though offering strong potential for creating highly-entangled states, the small size of optical cavities is a disadvantage when it comes to transferring quantum states over long distances, which is important for applications in quantum communications. For long distance entanglement, coupling of the qubits to reservoirs supporting propagating photonic modes, such as photonic crystals, is useful. In particularly, efficient transfer of entanglement between two quantum dots (QDs) placed inside a photonic crystal over a distance 100 times the wavelength of vacuum radiation has been theoretically predicted \cite{Yao09}. 

Despite the success of using all-dielectric photonic reservoirs for preparation and transfer of entanglement, the large size of such structures is a drawback for their use in nanophotonic integrated circuits. In order to overcome this difficulty, significant attention has been recently devoted to plasmonic waveguides as intermediaries for quantum state transfer. Nanoscale localization of plasmonics modes, as well as their relatively long propagation lengths, makes them good candidates for photonic circuits \cite{Tame13}. Efficient plasmon-mediated entanglement has been predicted for emitters coupled to a metal or metamaterial slab \cite{Xu11}, infinitely long V-shaped waveguides cut in a flat metal plane \cite{Gonzalez,Garcia}, infinitely long nanowires \cite{Dzsotjan,Garcia,Zheng13,Gonzalez14}, and arrays of metallic nanospheres \cite{Changhyoup}. Transfer of entanglement over distances far-exceeding the radiation wavelength was reported \cite{Garcia}. 

 \begin{figure}[h]
        \begin{center}
                \noindent
                \includegraphics[width=4.5in]{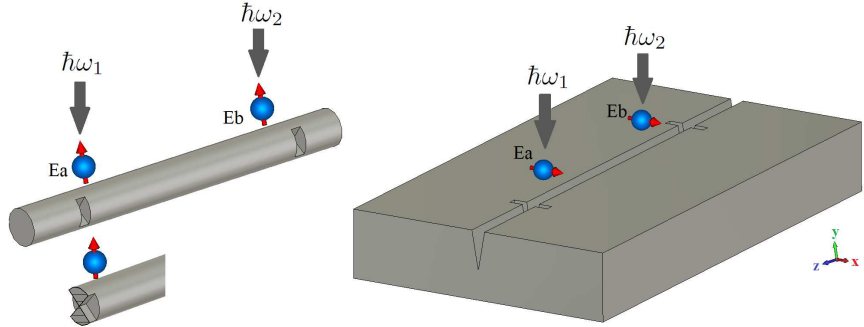}
                \caption{Two identical two-level emitters  (e.g., atoms or quantum dots) placed next to (a) a metal nanowire (insert shows geometry of the nanowire with coupling slots; the slot opening angle and width are $95^{\circ}$ and 15 nm, respectively) and (b) a V-shaped channel cut in a flat metal plane (if coupling slots are present, their length, width and depth are 70 nm, 15 nm, and 138 nm, respectively). We consider both infinite- and finite-length waveguides.}\label{Fig1}
        \end{center}
 \end{figure}

One can ask if it is possible to optimize the geometry of a plasmonic waveguide in order to achieve stronger plasmon mediated entanglement between qubits. Moreover, realistic waveguides are never infinite, and can have discontinuities, accidental or by design. Therefore, it is important to understand what role waveguide edges and inhomogeneities can play in entanglement transfer, which is the subject of this paper. We consider two quantum emitters (modeled as two-level systems) placed above 3D waveguides. Since there are a multitude of possible plasmonic waveguide geometries that one can envision, here we restrict our consideration to two waveguides that have been extensively studied in the literature, i.e. a metal nanowire and a V-shaped channel cut in a flat metal plane (see Fig. \ref{Fig1}). Regarding long-distance entanglement, these waveguide geometries have been studied assuming that the nanowire and groove are infinite in their axial directions \cite{Garcia, Gonzalez}. The main result of this work is to show that realistic finite-length nanowire and groove waveguides, with their associated discontinuities, play a crucial role in the engineering of highly entangled states. We demonstrate that proper positioning of the emitters with respect to the waveguide edges can lead to a significant increase in entanglement compared to the case of the emitter coupled to an infinite plasmonic waveguide. Moreover, even for the infinite-length case, discontinuities in the waveguides do not always play a detrimental role; an increase in entanglement compared to the unperturbed case can be achieved by introducing coupling slots into the structure (which aid in qubit-waveguide coupling). In the following we demonstrate the effect of coupling slots for both finite-length and infinite-length waveguides.

The paper is organized as follows. In Section \ref{Sec:Theory} we present the theory describing dissipation-induced entanglement between two emitters coupled to a plasmonic waveguide. We briefly introduce the concept of the Green dyadic (Sec. \ref{Sec:Dyadic}) as an entity rigorously describing the spatial and spectral distribution of plasmonic modes in the dissipative structured reservoir formed by the waveguide. We then present quantum master equations (Sec. \ref{Sec:transient}) that use the medium's Green function, which describe the evolution of the density matrix of two emitters, one of which is initially in its excited state, placed near the lossy waveguide. We solve these equations numerically and calculate the degree of entanglement induced by the plasmonic waveguide. The entanglement presented in Sec. \ref{Sec:transient} is a transient one as it decays with time due to depopulation of the emitters' excited states. In order to obtain a steady entangled state, the emitters depopulation has to be compensated by constant pumping with an external electromagnetic field, which is the subject of Sec. \ref{Sec:Steady}. Numerical results for entanglement are presented in Sec. \ref{Sec:Numer}. Transient entanglement between emitters coupled to a silver nanowire or a V-shaped channel waveguide is studied in Sec. \ref{Sec:Nanowire} and Sec. \ref{Sec:Groove}, respectively, and results for steady-state entanglement are presented in Sec. \ref{Sec:Steady_numer}.

\section{Theory}
\label{Sec:Theory}
In this section we introduce two concepts that are crucial for the theoretical description of the evolution of quantum systems coupled to lossy environments, namely, the dyadic Green function \cite{Welsch1, Trung} and the quantum master equation \cite{Dzsotjan,Garcia}. We use these concepts in order to describe behavior of two identical two-level emitters (qubits) placed above 3D plasmonic waveguides (see Fig. \ref{Fig1} for the waveguide geometries).  

\subsection{Dyadic Green function}
\label{Sec:Dyadic}
Even though the process of spontaneous decay is essentially non-classical, and thus requires a full quantum description, the media the emitter is coupled to can be rigorously incorporated into the quantum formalism through a classical quantity known as the dyadic Green function $\textbf{\textit{G}}$ \cite{Welsch1, Trung}. The electric field Green function is a 3-by-3 dyadic, each column of the dyadic being the electric field produced by a classic electric dipole polarized along the corresponding coordinate system axis. The Green dyadic satisfies
\begin{equation} \label{Eq:Green_dyad_eq}
\nabla \times \nabla \times \textbf{\textit{G}}(\textbf{r},\textbf{r}',\omega)-k_0^2 \varepsilon(\textbf{r},\omega) \textbf{\textit{G}}(\textbf{r},\textbf{r}', \omega) = k_0^2 \textbf{\textit{I}} \delta(\textbf{r}-\textbf{r}'),
\end{equation}
where $\mathbf{r},\mathbf{r'}$ are the observation and source point vectors, respectively, $ k_0 = \omega /c $ is the vacuum wavenumber, $\omega$ is angular frequency, $c$ is the speed of light in vacuum, $\varepsilon(\textbf{r},\omega)$ is relative permittivity, and $\textbf{\textit{I}}$ is the unit 3-by-3 dyadic.  The solution of \eqref{Eq:Green_dyad_eq} can be written as
\begin{equation}
\textbf{\textit{G}}(\textbf{r},\textbf{r}',\omega) = \textbf{\textit{G}}^{(0)}(\textbf{r},\textbf{r}',\omega) + \textbf{\textit{G}}^{(\mathrm{sc})}(\textbf{r},\textbf{r}',\omega),
\end{equation}
where
\begin{equation} \label{Eq: Green_dyad_free}
 \textbf{\textit{G}}^{(0)}(\textbf{r},\textbf{r}',\omega) = \left(k_0^2 + \nabla \otimes \nabla \right) \frac{e^{i k_0 |\textbf{r} - \textbf{r}'|}}{|\textbf{r} - \textbf{r}'|},
\end{equation}
is the free-space Green dyadic (i.e. the solution of \eqref{Eq:Green_dyad_eq} assuming $\varepsilon(\textbf{r},\omega) = 1$), $\otimes$ is the dyadic product, and $\textbf{\textit{G}}^{(\mathrm{sc})}(\textbf{r},\textbf{r}',\omega)$ is the ``scattered'' Green function, which is the homogeneous solution of \eqref{Eq:Green_dyad_eq}, accounting for the electric field scattered by the media. It should be noted that although the real part of $\textbf{\textit{G}}^{(0)}(\textbf{r},\textbf{r}',\omega)$ is singular when $\mathbf{r} = \mathbf{r}'$, the imaginary part is finite, with $\mathrm{Im} G_{jj}^{(0)}(\textbf{r},\textbf{r},\omega) = { k_0^3}/{6\pi}$, $j = x,y,z$. In contrast to the free-space Green dyadic, the scattered Green dyadic $\textbf{\textit{G}}^{(\mathrm{sc})}(\textbf{r},\textbf{r}',\omega)$ does not have any singularities. However, it depends both on the geometry and permittivity of objects filling the space, and thus does not have a closed analytic form in the general case. In order to calculate the Green dyadic $\textbf{\textit{G}}(\textbf{r},\textbf{r}',\omega)$ for the metallic plasmonic waveguides presented in Fig.~\ref{Fig1} we numerically solved Maxwell equations \eqref{Eq:Green_dyad_eq} using a commercial finite-difference time-domain method (FDTD) from Lumerical solutions \cite{Lumerical2}. 

The importance of the classical Green function is that it leads to the dissipative decay rate and coherent coupling terms
\begin{align}
\Gamma_{\alpha\beta}(\omega_{\alpha}) & = \frac{2 }{\varepsilon_0 \hbar }  \mathrm{Im}\,  \mathbf{d} \cdot \textbf{\textit{G}}(\mathbf{r}_{\alpha}, \mathbf{r}_{\beta},\omega_{\alpha}) \cdot \mathbf{d}, \label{Eq:decay_rate}\\
g_{\alpha\beta}(\omega_{\alpha}) & = \frac{1}{\varepsilon_0 \hbar}  \mathrm{Re}\,  \mathbf{d} \cdot \textbf{\textit{G}}(\mathbf{r}_{\alpha}, \mathbf{r}_{\beta},\omega_{\alpha}) \cdot \mathbf{d}.
\end{align}
Here $\mathbf{d}$ is the emitter transition dipole moment, $\Gamma_{\alpha\alpha}$ and $\Gamma_{\alpha\beta}(\alpha\neq \beta)$ are the decay rates of the emitter $\alpha$ due to its interaction with the reservoir (which includes the plasmonic system), and plasmon mediated interactions with the emitter $\beta$, respectively. When the emitter itself is in a lossless region of space (such as the space above the waveguide)$, \mathrm{Im}\textbf{\textit{G}}(\mathbf{r}_{\alpha}, \mathbf{r}_{\alpha},\omega_{\alpha})$ does not have any singularity and thus the decay rate is always finite. The emitters' transition frequency shift induced by dipole-dipole coupling is given by $g_{\alpha\beta}(\alpha \neq \beta) $. We assume that the emitter transition frequency $\omega_{\alpha}$ already accounts for the photonic Lamb shift which is defined by $g_{\alpha\alpha}$. In \cite{Garcia} it was shown that for an infinite waveguide the best entanglement was obtained when $\Gamma_{\alpha\beta}$ was large and $g_{\alpha\beta}$ was small (forming the dissipative regime). For the finite waveguide case this classification does not hold exactly, which is discussed below (e.g., see Fig. \ref{Fig2_2}).

In the following we consider two identical emitters located near one of the plasmonic waveguides shown in Fig. \ref{Fig1} and assume that the waveguide material is silver, which ia relatively low loss metal. We choose the dipole transition frequency of the emitters to be $ \omega_a/2\pi=\omega_b/2\pi=500 $ THz (emission wavelength $ \lambda = 600$ nm). The relative permittivity of silver at the dipole transition frequency is $-13.9 + i 0.92$ \cite{Palik}, and we fully account for the dispersive properties of the metal. For the nanowire waveguide we choose the radius to be $R = 35$ nm, such that the surface plasmon wavelength in the nanowire is $\lambda_{\mathrm{spp}} = 425$ nm, and the propagation length is $l=1.7~\mathrm{\mu} m$, calculated analytically \cite{Novotny}. For the groove waveguide, the depth of the groove is $ 138 $ nm and its opening angle is $ 20^{\circ} $; these dimensions were chosen in \cite{Garcia} to yield almost identical plasmonic wavelengths and propagation lengths for the rod and groove structures (for the groove $ \lambda_{\mathrm{spp}}= 423 $ nm and $l=1.7~\mathrm{\mu} m$), facilitating comparison between the waveguides.

The total emitter decay rate can be expressed as $\Gamma_{aa} = \Gamma_{\mathrm{rad}} + \Gamma_{\mathrm{Joule}} + \Gamma_{\mathrm{pl}}$, in terms of the following emitter decay channels: (a) free-space radiation, $ \Gamma_{\mathrm{rad}}$, (b) Joule losses in the metal, $\Gamma_{\mathrm{Joule}}$, and (c) excitation of surface plasmons, $\Gamma_{\mathrm{pl}}$. Since radiative decay is not helpful for long-distance entanglement, and Ohmic decay is obviously deleterious, it is clear that we want to maximize the plasmon decay channel, which enhances qubit-qubit interaction. This requires careful positioning of the emitter above the nanowire surface; if the emitter is too close to the nanowire surface, Joule losses dominate the emitter decay (quenching). On the other hand, if the emitter is too far away than it mostly radiates into free space. This problem was studied in \cite{Garcia} where it was demonstrated that at $\lambda=600$ nm the optimum distance between the surface of an $R=35$ nm infinitely-long silver nanowire and a tuned emitter is 20 nm (which maximizes $\Gamma_{\mathrm{pl}}$), with dipole moment perpendicular to the wire, and for the groove structure the optimum emitter height is 12 nm above the surface of the metal, positioned over the central line of the groove and polarized horizontally. 

For the finite nanowire the problem is more complicated and a unique definition of $\Gamma_{\mathrm{pl}}$ is difficult (since plasmons reflect and refract from the nanowire edges). We expect that the optimum distance should not be considerably different from the infinite-nanowire case, and we here choose the separation between the emitter and waveguide to be the same as the optimum height for the infinite structure, 20 nm for the nanowire and 12 nm for the groove. Although not shown here, slight improvement in entanglement can be obtained by adjusting the qubit height by a few nm. 

\begin{figure}[h]
        \begin{center}
                \noindent
                \includegraphics[width=5in]{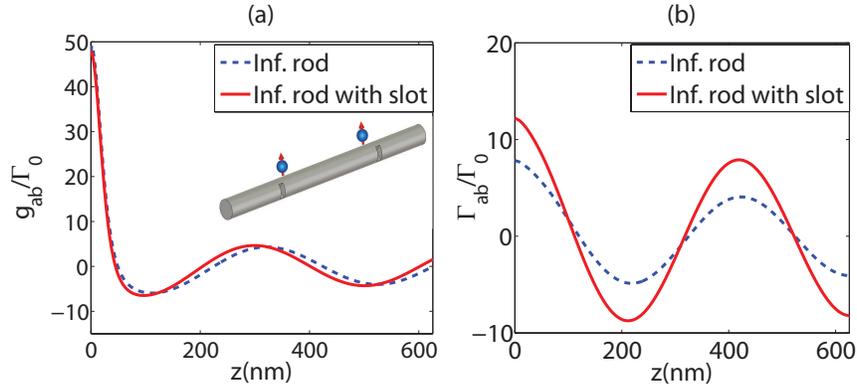}
                \caption{Normalized coupling strength $g_{ab}$ and decay rate $\Gamma_{ab}$ as a function of distance between emitters above an infinite-length silver nanowire with and without coupling slots (based on scattered Green function). The position of one emitter is fixed at $z=0$ and the second emitter position is changing with $z$. Slot are located at $z=-40$ nm and $z= 1.5\lambda_{\mathrm{spp}}+40$ nm $=$ 677.5 nm.}\label{Fig2_1}
        \end{center}
\end{figure}

Although a primary aim of this work is to examine the effect of a finite plasmonic waveguide structure on entanglement, we first show that introducing discontinuities on an infinite waveguide can enhance qubit-waveguide coupling (i.e., in the following we refer to these discontinuities as coupling slots), thereby enhancing entanglement. Figure \ref{Fig2_1} shows $ \Gamma_{ab}/\Gamma_0 $ and $ g_{ab}/\Gamma_0 $ for emitters placed above an infinite-length nanowire, where $z$ indicates the lateral separation between emitters. One emitter is located at $z=0$, and the position of the other emitter varies from $z=0$ to $z=1.5\lambda_{\mathrm{spp}}=637.5$ nm. Note that here, and in Fig. \ref{Fig2_2} below, we plot the rates based on the scattered Green function; this term is dominate over the vacuum Green function for all cases of interest (otherwise the plasmonic system would not affect the qubits much, such as, e.g., if the height of the qubits was increased considerably), and it avoids the singularity in $g_{aa}$ associated with the vacuum Lamb shift. However, in computing entanglement we use the full (scattered plus vacuum) Green function.

Oscillatory behavior of $ \Gamma_{ab}$ and $g_{ab}$ as a function of qubit separation distance is evident in Fig. \ref{Fig2_1}, with the period of oscillation being approximately equal to the plasmon wavelength. Moreover, the positions of the $ \Gamma_{ab} $ maxima correspond to positions of $g_{ab}$ minima. Thus coherent and dissipative regimes become dominant at different separations between emitters \cite{Garcia}. For the case of a nanowire with coupling slots, the slots are fixed at $z=-40$ nm and $z= 1.5\lambda_{\mathrm{spp}}+40$ nm$=$677.5 nm (one slot is to the right of the right emitter, and the other to the left of the left emitter; $z$ refers to emitter-emitter separation). Coupling slots can lead to enhancement of entanglement between the emitters by increasing the dissipative decay rate at the emitter positions, as shown in the figure (for the effect of coupling slots on entanglement, see Fig. \ref{Fig4} below). By adding judiciously-chosen coupling slots to the infinitely long nanowire we are increasing emitter-waveguide coupling, similar to the improved field-plasmon coupling in the case of a grating, and thereby increasing emitter decay rates $ \Gamma_{aa} $ and $\Gamma_{ab}$ into the plasmon channel. It should be noted that coupling slots seem to affect only the dissipative regime, as the slot induced modification of the coherent exchange rate seems to be insignificant (Fig. \ref{Fig2_1}a). The normalization constant
\begin{equation}
\Gamma_{0} = \frac{2 }{\varepsilon_0 \hbar }  \mathrm{Im}\,  \mathbf{d} \cdot \textbf{\textit{G}}^{(0)}(\mathbf{r}_{\alpha}, \mathbf{r}_{\alpha},\omega_{\alpha}) \cdot \mathbf{d} =  \frac{\omega_0^3 d^2}{3\pi\varepsilon_0 \hbar c^3}
\end{equation}
is the decay rate of the emitter in free-space. It should be noted that the ratios $ \Gamma_{ab}/\Gamma_0 $ and $ g_{ab}/\Gamma_0 $ do not depend on the emitter dipole moment $\mathbf{d}$, since we are in the weak coupling regime. 

It is likely that further improvement in emitter-plasmon coupling (and, subsequently, entanglement) can be made by optimizing the slot geometry. In fact, qubit entanglement depends on many parameters even if the dipole transition frequency and nanowire diameter are fixed. In the case of an infinitely-long nanowire these parameters are the distance between emitters and the nanowire surface, separation between emitters, and position and geometry of the coupling slots. In the case of a finite-length nanowire we also have to consider the nanowire length, and positions of the emitters and coupling slots relative to the nanowire ends. In theory, optimal values of these parameters could be found though multi-variable optimization. However, in this work we make no attempt to find optimal values; we merely show that reasonably-positioned qubits and coupling slots can improve entanglement. 

\begin{figure}[h]
        \begin{center}
                \noindent
                \includegraphics[width=5in]{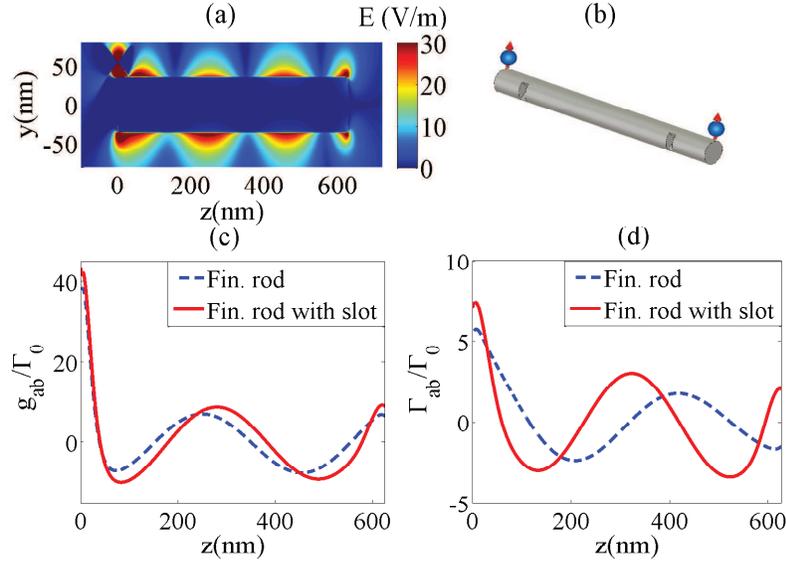}
                \caption{(a) Absolute value of plasmon electric field intensity on a finite nanowire of length $ 1.5\lambda_{\mathrm{spp}} $. The plasmon mode is excited by the emitter positioned at $z = 0$ nm at a distance $ 20 $ nm above the nanowire surface. (b) geometry, (c)-(d) Normalized coupling strength $g_{ab}$ and decay rate $\Gamma_{ab}$ as a function of a distance between emitters above the nanowire with and without coupling slots (based on scattered Green function). Slot are located at $z=27$ nm and $z= 1.5\lambda_{\mathrm{spp}}-27$ nm=610.5 nm.}\label{Fig2_2}
        \end{center}
\end{figure}

For the finite-length nanowire we choose the nanowire length to be $1.5\lambda_{\mathrm{spp}}=637.5$ nm.  We assume one emitter is located above one end of the nanowire ($ z=0 $ nm). If coupling slots are present, they are placed 27 nm away from each end of the nanowire (this location leads to enhanced $ \Gamma_{ab}$ when qubits are at the nanowire ends). The spatial distribution of the electric field intensity of the plasmon mode is shown in Fig. \ref{Fig2_2}a. In order to calculate this distribution we assume a classical electric dipole at the first emitter's position. We can see that the positioned emitters experience maximum intensity of the plasmon electric field.

Figures \ref{Fig2_2}(b)-(c) show the dissipative and coherent rates as a function of emitter-emitter separation for the finite-length nanowire. We can see that due to plasmon reflections from the nanowire edges, positions of maximums of $g_{ab}$ and $\Gamma_{ab}$ almost coincide with each other. Thus we can not clearly separate the dissipative and coherent regimes in the emitter dynamics anymore, unlike for the infinite-waveguide case. For emitters positioned over the nanowire ends, the addition of coupling slots leads to an increase of coupling between emitters and surface plasmons. Moreover, in this case we observe a shift in the position of the maximums of the emitter coupling rates (which does not occur for the infinite-length structure), as the slots change the effective length of the nanowire. Although not showm, similar results and conclusions are obtained for the groove waveguide.

\subsection{Transient entanglement between two plasmon-coupled qubits}
\label{Sec:transient}
Let us consider two identical two-level emitters placed above the plasmonic waveguide (see Fig.~\ref{Fig1}). Evolution of the system density matrix $\rho$ is described by the Von Neumann equation, 
\begin{equation} \label{Eq:Neumann}
\partial_t \rho = -\frac{i}{\hbar}\left[H,\rho \right],
\end{equation}
where $H$ is the system Hamiltonian, including emitters and plasmonic reservoir degrees of freedom, as well as coupling between them \cite{Dzsotjan,Garcia}, 
\begin{align}
H & =  \int d \textbf{r} \int_{0}^{+\infty} d\omega_{\lambda} \, \hbar \omega_{\lambda} \textbf{b}^{\dagger}{(\textbf{r},\omega_{\lambda})}\cdot \textbf{b}{(\textbf{r},\omega_{\lambda})}
+\displaystyle\sum_{\alpha=a,b} \hbar \omega_{\alpha}  \sigma^{\dagger}_{\alpha} \sigma_{\alpha} \notag \\
& - \displaystyle\sum_{\alpha=a,b}(\sigma_{\alpha} + \sigma^{\dagger}_{\alpha}) \, \textbf{d} \cdot {\textbf{E}}{(\textbf{r}_\alpha)}.
\end{align}
Here the index $\alpha = a,b$ indicates the emitters, $\omega_{\alpha}$ and $\textbf{d}$ are the frequency and transition dipole moment between the emitter ground $\left|g\right>_{\alpha}$ and excited $\left|e\right>_{\alpha}$ states, $\sigma_{\alpha} = \left|g\right>_{\alpha} \left<e\right|_{\alpha} $ and $\sigma_{\alpha}^{\dagger} = \left|e\right>_{\alpha} \left<g\right|_{\alpha}$ are the emitter lowering and raising operators, respectively. The operator of the electromagnetic field in the presence of the plasmonic reservoir is \cite{Welsch1, Trung}
\begin{equation}
{\textbf{E}}{(\textbf{r})}=i \sqrt{ \frac{\hbar}{\pi\varepsilon_0}}\int_{0}^{+\infty} d\omega_{\lambda} \int\sqrt{\mathrm{Im}(\varepsilon(\textbf{r}';\omega_{\lambda})} \textbf{\textit{G}}{(\textbf{r},\textbf{r}',\omega_{\lambda})} \cdot \textbf{b}{(\textbf{r}',\omega_{\lambda})} d\textbf{r}' 
+\mathrm{H.c.},
\end{equation}
where $\textbf{b}{(\textbf{r}',\omega_{\lambda})}$ are operators of elementary excitations of the reservoir, satisfying bosonic commutation relations $\left[b_i{(\textbf{r},\omega_{\lambda})},b_j^{\dagger}{(\textbf{r}',\omega_{\lambda'})}\right] = \delta_{ij} \delta(\omega_{\lambda}-\omega_{\lambda'}) \delta(\textbf{r}-\textbf{r}')$ and $\left[b_i{(\textbf{r},\omega_{\lambda})},b_j{(\textbf{r}',\omega_{\lambda'})}\right] = 0$. 

Tracing out the electromagnetic field degrees of freedom in \eqref{Eq:Neumann} and using the Born, Markov and rotating wave approximations, we obtain master equations for the reduced $4\times4$ density matrix as
\begin{align} \label{Eq:master}
\partial_t \rho_s(t)  = \mathcal{K} \rho_s(t),
\end{align}
where
\begin{align} 
\mathcal{K} \rho_s & = \sum_{\alpha,\beta = a,b} \frac{\Gamma_{\alpha \beta}(\omega_{\alpha})}{2} \left( 2 \sigma_{\alpha}\rho_s \sigma_{\beta}^{\dagger} -  \sigma_{\alpha}^{\dagger} \sigma_{\beta} \rho_s - \rho_s \sigma_{\alpha}^{\dagger}  \sigma_{\beta} \right) \notag \\
&  +  i \left[\left(g_{ab}(\omega_{\alpha})\hat{\sigma}_a^{\dagger} \hat{\sigma}_b + g_{ba}(\omega_{\alpha})\hat{\sigma}_b^{\dagger} \hat{\sigma}_a \right) ,\tilde{\rho}_s \right],
\end{align}
accounts for the interaction between the emitters and the plasmonic reservoir. It should be noted that Eq. \eqref{Eq:master} accounts for the emitter decay due to its interaction with the electromagnetic field of the plasmonic reservoir. However, depending on the nature of the emitter, there can be other non-radiative channels for emitter decay. In particular, in the case of QDs, electron-phonon interaction leads to non-radiative quenching of emitter luminescence. In order to account for these processes, we introduce pure dephasing phenomenologically, i.e.,
\begin{align} \label{Eq:master_deph}
\partial_t \rho_s(t)  = \mathcal{K} \rho_s(t) + \mathcal{L} \rho_s(t),
\end{align}
where 
\begin{align}
\mathcal{L} \rho_s = \sum_{\alpha = a,b} \frac{\gamma_{\alpha}(\omega_{\alpha})}{2} \left( 2 \sigma_{\alpha}\rho_s \sigma_{\alpha}^{\dagger} -  \sigma_{\alpha}^{\dagger} \sigma_{\alpha} \rho_s - \rho_s \sigma_{\alpha}^{\dagger}  \sigma_{\alpha} \right), 
\end{align}
and where $\gamma_{\alpha}$ is a phenomenological pure-dephasing rate. We show later that in the case of a typical QDs above a plasmonic waveguide, the dephasing decay rate is generally small compared to the radiative decay rate and thus it's contribution to QD decay dynamics is insubstantial. Unless specified otherwise (e.g., Fig. \ref{Fig6}), we assume it to be equal to zero.  

Equations \eqref{Eq:master_deph} can be solved analytically in the basis $\left|3\right> =  \left|e_a,e_b\right>$, $\left|0\right> =  \left|g_a,g_b\right>$, $
\left|\pm\right> = \frac{1}{\sqrt{2}} \left( \left|e_a,g_b\right> \pm \left|g_a,e_b\right> \right)$. Particularly, if initially the only non-zero elements of the density matrix are $\rho_{++}$, $\rho_{--}$, $\rho_{+-}$, $\rho_{-+}$, than the system of master equations is reduced to
\begin{align} \label{Eq:master_tr1}
&\frac{\partial \rho_{++}(t)}{\partial t} = - \left( \Gamma_{aa}'+\Gamma_{ab}\right) \rho_{++}(t), \quad \Rightarrow \quad \rho_{++}(t) = \rho_{++}(0) \, e^{-(\Gamma_{aa}'+\Gamma_{ab})t}, \\
&\frac{\partial \rho_{--}(t)}{\partial t} =  - \left(\Gamma_{aa}' - \Gamma_{ab}\right) \rho_{--}(t), \quad \Rightarrow \quad \rho_{--}(t) = \rho_{--}(0) \, e^{-(\Gamma_{aa}'-\Gamma_{ab})t}, \\
&\frac{\partial \rho_{+-}(t)}{\partial t} = - \left(\Gamma_{aa}' - 2 i g_{ab} \right) \rho_{+-}(t), \quad \Rightarrow \quad \rho_{+-}(t) = \rho_{+-}(0) \, e^{-(\Gamma_{aa}'-2 i g_{ab})t},  \label{Eq:master_tr3}
\end{align}
where $\Gamma_{aa}' = \Gamma_{aa} + \gamma_a$. 
This is the case when only one of the emitters is initially excited, such that $\rho_{++}(0) = \rho_{+-}(0) = \rho_{-+}(0) = \rho_{--}(0) = 1/2$. 

In order to characterize entanglement between two emitters we use the concept of concurrence \cite{Wooters,Aolita}, which is defined as 
\begin{equation} \label{Eq:concur_general}
C = \mathrm{max}(0,\sqrt{u_1} - \sqrt{u_2} - \sqrt{u_3} - \sqrt{u_4}),
\end{equation}
where $u_i$ are arranged in descending order of the eigenvalues of the matrix $\rho_s \tilde{\rho}_s$, where $\tilde{\rho}_s = \sigma_y \otimes \sigma_y \rho_s^{\star} \sigma_y \otimes \sigma_y$ is the spin-flip density matrix with $\sigma_y$ being the Pauli matrix. Concurrence may vary in the range between 0 (unentangled state) and 1 (completely entangled), such that values between 0 and 1 correspond to different degrees of entanglement.

If the only non-zero elements of the density matrix are defined by Eqs. \eqref{Eq:master_tr1}-\eqref{Eq:master_tr3}, then concurrence is given by simplified expression
\begin{align} \label{Eq:concur}
C(t) & = \sqrt{\left[\rho_{++} - \rho_{--}\right]^2 + 4 \mathrm{Im} [\rho_{+-}]^2} =  \frac{e^{-\Gamma_{aa} t}}{2}  \sqrt{\left[e^{-\Gamma_{ab}t} - e^{\Gamma_{ab}t}\right]^2 + 4 \sin^2(2  g_{ab}t)},
\end{align}
where we again assumed that only one of the emitters is initially in the excited state.

\subsection{Steady-state entanglement between two qubits}
\label{Sec:Steady}
As one can see from Eq. \eqref{Eq:concur}, the entanglement between two emitters above a plasmonic waveguide decays with time. In order to obtain a steady entangled state we need to compensate for the depopulation of the emitters' excited states (via coupling to the environment) by pumping with an external laser field. Here we restrict our consideration to the case when the external pumps are modeled by classical monochromatic waves of frequency $\omega_L$, $\mathbf{E}_{\alpha} = \mathbf{E}_{0\alpha} e^{-i\omega_{L} t} + \text{c.c.}$, and assume that each of the emitters can be pumped individually (i.e., using focused beams). In this case, the master equation takes the form
\begin{align} \label{Eq:master_pump}
\partial_t \rho_s(t)  = -(i/\hbar) \left[V,\rho_s(t)\right] + \mathcal{K} \rho_s(t) + \mathcal{L} \rho_s(t) ,
\end{align}
where
\begin{equation}
V = - \hbar \sum_{\alpha = a,b} \left( \Omega_{\alpha}  e^{- i \Delta_{\alpha} t} \sigma_{\alpha}^{\dagger} + \Omega_{\alpha}^* e^{i \Delta_{\alpha} t} \sigma_{\alpha} \right)
\end{equation}
accounts for the interaction between the classical pump field and qubits, $\Omega_{\alpha} = \mathbf{d} \cdot \mathbf{E}_{0\alpha} / \hbar$ is the effective Rabi frequency of the pump, and $\Delta_{\alpha} = \omega_{\alpha} - \omega_L$ is a detuning parameter. In the basis $\left|e_a,e_b\right>$, $\left|e_a,g_b\right>$, $\left|e_b,g_a\right>$, $\left|g_a,g_b\right>$ we obtain a system of 16 coupled differential equations for the density matrix, which we solve numerically. For the pumped case, Eq. \eqref{Eq:concur_general} has to be used to calculate concurrence.

\section{Numerical results}
\label{Sec:Numer}

In order to calculate the transient concurrence we use the analytic equation \eqref{Eq:concur}. Steady-state concurrence is calculated using \eqref{Eq:concur_general} and density matrix elements are obtained by solving numerically the system of 16 differential equations. Unless specified otherwise (as in Fig. \ref{Fig6}), the pure dephasing rate of the emitters, $\gamma_a$, is equal to zero.

\subsection{Transient entanglement mediated by plasmons in a metallic nanowire}
\label{Sec:Nanowire}

\begin{figure}[h!]
	\begin{center}
		\noindent
		\includegraphics[width=4.5in]{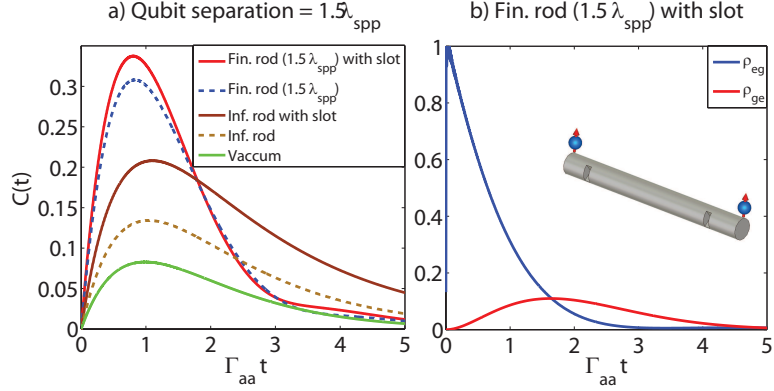}
		\caption{ (a) Time dependence of transient concurrence between two emitters laterally separated by $ 1.5\lambda_{\mathrm{spp}}=637.5 $nm. We assume that the emitters are in vacuum (solid green line), near an infinite wire with (dashed brown line) and without (solid brown line) coupling slots, near a finite wire of length of $1.5\lambda_{\mathrm{spp}}$ with (solid red) and without (dashed blue line) coupling slots. In the case of the finite wire, qubits are placed above the nanowire ends, and slots are located 27 nm from each end. The normalization constant $\Gamma_{aa}$ is the decay rate of the emitter (Eq. \eqref{Eq:decay_rate}) for the given geometry. (b) Time dependence of the population of the emitter initially in the excited, $\rho_{eg}$, and ground, $\rho_{ge}$, states, for the case of a finite nanowire with coupling slots. }\label{Fig4}
	\end{center}
\end{figure}

It was reported in Ref. \cite{Garcia} that entanglement between emitters coupled to an infinite length plasmonic nanowire exceeds that between emitters in vacuum. As one can see from Fig. \ref{Fig4}a entanglement can be improved even further by introducing discontinuities (coupling slots) in the nanowire, as described in Sec. \ref{Sec:Dyadic}. In particular, using an infinite nanowire with coupling slots (see Fig. \ref{Fig1} and Sec. \ref{Sec:Dyadic} for the details of the slot geometry) can considerably increase concurrence compared to the case with no slots, due to improved emitter-nanowire coupling. 

Fig. \ref{Fig4}a also shows results for the concurrence of two qubits near a finite-length nanowire, with length equal to $ 1.5\lambda_{\mathrm{spp}} $, forming a Fabry-P\'{e}rot (FP) cavity, with the qubits positioned above the ends of the wire. It can be seen that by coupling into the FP nanowire resonances, we can considerably improve entanglement compared to the infinite-length case. Figure \ref{Fig4}b shows evolution of the emitters excited state population for the case of a finite nanowire with coupling slots. Here $ \rho_{eg} $ is the probability of the first qubit to be in the excited state and the second qubit to be in ground state, and $ \rho_{ge} $ is similarly defined. The initial state is $\rho_{eg} = 1$. We can see that decay of the emitter initially in the excited state leads to excitations of the emitter initially in the ground state, creating a transient entangled state, which is decaying with time.

\subsection{Transient entanglement mediated by V-shaped groove waveguide}
\label{Sec:Groove}

\begin{figure}[h!]
	\begin{center}
		\noindent
		\includegraphics[width=4.5in]{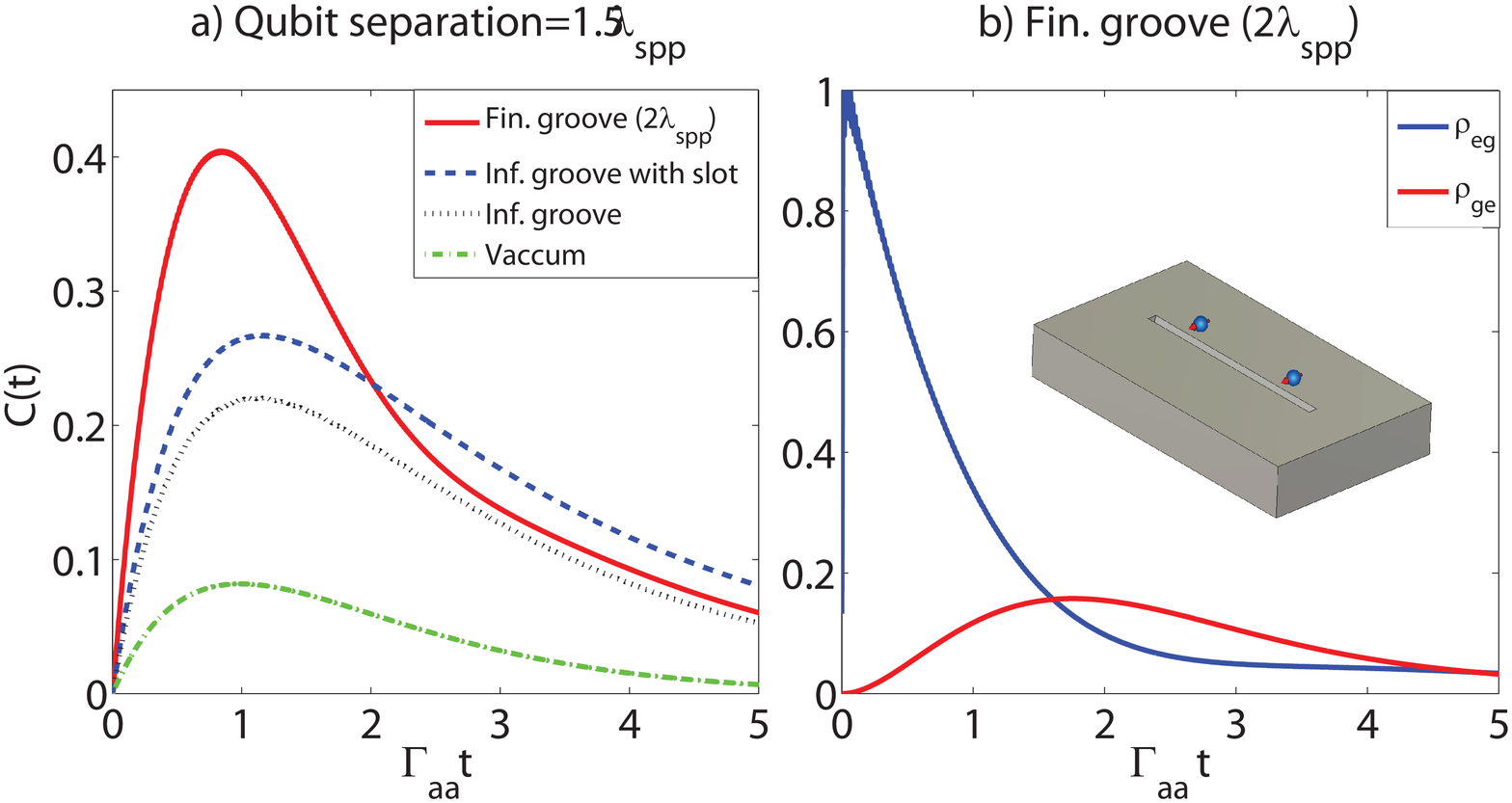}
		\caption{ (a) Concurrence of two qubits separated by $1.5\lambda_{\mathrm{spp}}$. Vacuum case (dashed green line), infinite groove (dotted black line), infinite groove with coupling slots (dashed blue line) $40$ nm laterally away from the qubits (outside the space between them), and finite groove of length $ 2\lambda_{\mathrm{spp}}$ (solid red line). (b) Time dependence of the population of the emitter initially in the excited, $\rho_{eg}$, and ground, $\rho_{ge}$, states. }\label{Fig5}
	\end{center}
\end{figure}

\begin{figure}[h!]
	\begin{center}
		\noindent
		\includegraphics[width=4in]{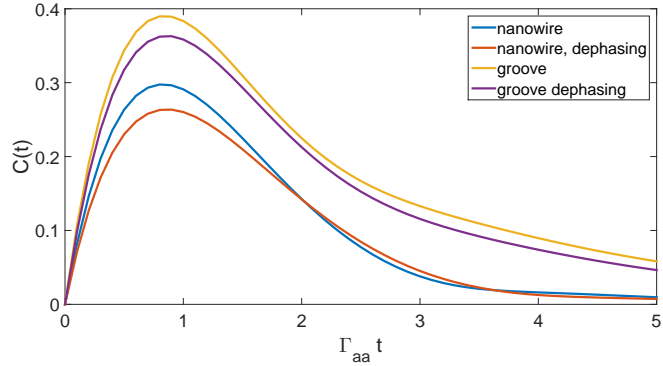}
		\caption{ Time dependence of the concurrence between two qubits including the effect of pure dephasing. Qubits are placed either above the ends of a finite-length nanowire (Fig. \ref{Fig4}a) or finite-length groove (Fig. \ref{Fig5}a)}\label{Fig6}
	\end{center}
\end{figure}

The concurrence of two qubits separated by a distance $ 1.5 \lambda_{\mathrm{spp}}=634.5 $ nm is shown in Fig. \ref{Fig5}a for an infinite groove, an infinite groove with coupling slots (slots are positioned as in the nanowire case), and a $ 2\lambda_{\mathrm{spp}}$ groove. Although the finite nanowire considered above had length $ 1.5\lambda_{\mathrm{spp}}$, for the finite groove the length is $ 2\lambda_{\mathrm{spp}}$ because for the groove the modal field is not strong at the groove ends (although not shown, the modal field distribution in the groove is more cavity-like, with a node at the groove end-wall), and so a $ 2\lambda_{\mathrm{spp}}$ groove allows the qubits to be separated by $ 1.5\lambda_{\mathrm{spp}}$ and also to be at the modal antinodes. Results for the finite-length groove are shown without slots since for this case coupling slots did not improve concurrence considerably; for the nanowire the modal field is strong at the wire ends (Fig. \ref{Fig2_2}), but it is not exactly a modal antinode because of strong end-diffraction. In this case, coupling slots serve to shift the mode pattern slightly, enhancing qubit-plasmon coupling. If the qubits are positioned exactly at a modal antinode, which is better defined away from the wire ends, coupling slots have a weaker effect on improving concurrence. It can be seen that for the infinite groove, the addition of coupling slots leads to improved concurrence, and the FP resonances of the finite-length groove leads to further improvement. As reported in \cite{Garcia}, for the chosen geometry the groove waveguide leads to stronger entanglement compared to the nanowire (although this conclusion is likely dependent on metal absorption and geometry). The population dynamics of the qubits are also shown for the finite groove case in Fig. \ref{Fig5}b.

In order to study the effect of pure dephasing, we considered a QD with dipole moment 30 D as our model system. It was reported \cite{Angelatos2014} that the dephasing rate, $\gamma_a$ of the QD is equal to 1 $\mu$eV. We considered emitters placed either above a finite nanowire (Fig. \ref{Fig4}a) or a finite groove (Fig. \ref{Fig5}a) without coupling slots. The emitter coupling rates for the case of the nanowire are $\Gamma_{aa} = 6.5 \mu$eV, $\Gamma_{ab} = - 1.2 \mu$eV, $g_{ab} = 2.85\mu$eV, while for the case of the groove waveguide  $\Gamma_{aa} = 11.38 \mu$eV, $\Gamma_{ab} = - 6.48 \mu$eV, $g_{ab} = 5.8\mu$eV. We can see that in general the pure dephasing decay rate is smaller than the radiative decay rate. Even though it affects entanglement detrimentally, the decrease in concurrence is rather small (see Fig. \ref{Fig6}). Thus we otherwise ignore pure dephasing in our results, although the validity of this assumption depends on the QD system under consideration and the temperature.

\subsection{Steady state entanglement under external pumping}
\label{Sec:Steady_numer}

Up to this point we have considered spontaneous (or vacuum-induced) decay which leads to transient entanglement between two coupled qubits. Transient entanglement does not persist due to depopulation of the excited state of the qubits. Thus, in order to obtain a steady entangled state, depopulation has to be compensated by pumping qubits with an external laser pump in resonance with the frequency of the qubits' dipole transition. We assume the pump interacts only with the qubits. In this case, the concurrence reaches a final, steady state (SS) value. 

\begin{figure}[h!]
	\begin{center}
		\noindent
		\includegraphics[width=5.5in]{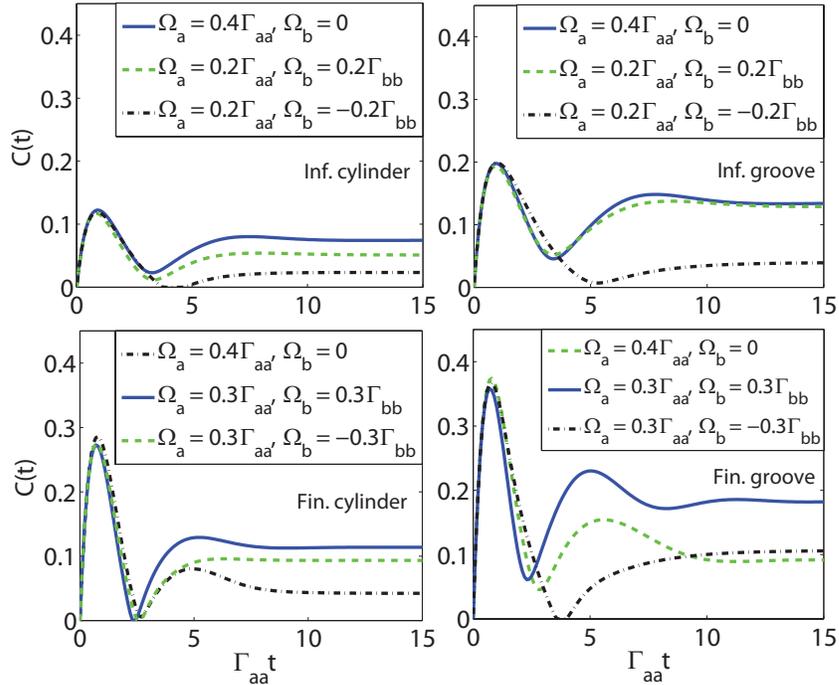}
		\caption{Time dependence of the concurrence between two qubits pumped by external electromagnetic fields. Qubits are placed above either a nanowire (left panel) or groove (right panel). Qubit separation is $ 1.5\lambda_{\mathrm{spp}} $. The system geometry is the same as in Figs. \ref{Fig4},\ref{Fig5}.}\label{Fig7}
	\end{center}
\end{figure}

Figure \ref{Fig7} shows the externally driven concurrence of qubits in the presence of a finite nanowire with length $ 1.5\lambda_{\mathrm{spp}} $ with qubits located at the nanowire ends (the same geometry as in Fig. \ref{Fig4}), a finite groove with length $ 2\lambda_{\mathrm{spp}} $, qubits located symmetrically above its surface separated by $ 1.5\lambda_{\mathrm{spp}} $ (the same geometry as in Fig. \ref{Fig5}), and the infinite-length cases with the same qubit spacings as in the finite case. Three different coherent pumps differing in the relative phase of the laser field on qubit 1 and 2 have been considered: symmetric pumping indicates identical Rabi frequencies, $\Omega_1=\Omega_2 $, antisymmetric pumping corresponds to  $ \Omega_1=-\Omega_2 $ and asymmetric pumping means $ \Omega_1\neq0,~  \Omega_2=0 $.

From Fig. \ref{Fig7}, it is clear that the transient portion of the concurrence is intensity independent, and that the infinite groove has a higher transient peak compared to the infinite nanowire for all pumping regimes. Numerical experimentation shows that larger values of $ g_{ab} $ and $ \Gamma_{ab} $ at the positions of the qubits lead to larger transient concurrence peaks. For the geometries studied, these mutual rates are larger for the infinite groove than for the infinite nanowire, which facilitates photon exchange between the qubits for the groove guide. We can also see that the main advantage of a finite waveguide relative to an infinite one is significant improvement in the transient concurrence peak, with more modest improvement in the final, steady-state value.

\begin{figure}[h!]
	\begin{center}
		\noindent
		\includegraphics[width=5.5in]{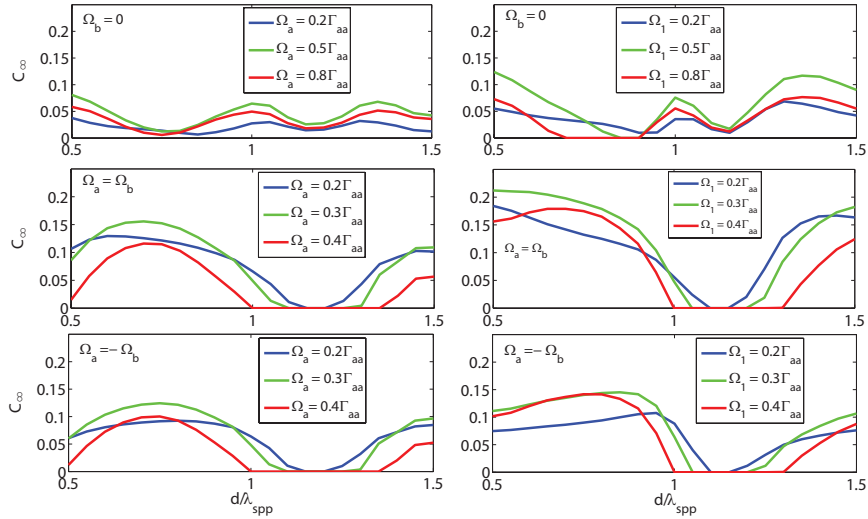}
		\caption{Steady state concurrence, $ C_{\infty} $, as function of qubit separation normalized by the modal wavelength, $ \lambda_{\mathrm{spp}}  $. Qubits are placed either above a finite nanowire (right) or a finite groove (left). The geometry is the same as in Figs. \ref{Fig4},\ref{Fig5}. Asymmetric, antisymmetric and symmetric pumping have been considered.}\label{Fig8}
	\end{center}
\end{figure} 

Figs. \ref{Fig8} shows the SS value of concurrence, $ C_{\infty} $, corresponding to different types of pumping as a function of qubit separation normalized to the plasmon wavelength, $ \lambda_{\mathrm{spp}}  $ for the finite nanowire (left panels) and finite groove (right panels). Numerical experimentation indicates that for steady state concurrence the preferred case is when we have large $ \Gamma_{ab} $ and small $ g_{ab} $, in agreement with \cite{Garcia} for the case of infinite waveguides. Therefore, the dissipative regime leads to larger transient entanglement peaks for infinite-length waveguides, and larger values of $ C_{\infty} $ in general, and it is only for transient entanglement on finite-length waveguides that the dissipative and coherent regimes becomes similar. Here, we can have steady state concurrence close to the peak of the transient value for relatively weak pumping. For steady state concurrence, large $ g_{ab} $ and $ \Gamma_{ab} $ are less ideal, because of the tendency to decrease SS concurrence. Furthermore, pump strength can not be too large otherwise the qubits will interact mostly with the pump and become decoupled from each other. Ideally, the pump should be strong enough to keep the system interacting, but small enough for the qubit interaction to dominate the dynamics. As made clear in Fig. \ref{Fig8}, pumping one qubit strongly (asymmetric pumping) is possible, but pumping both qubits strongly (symmetric or antisymmetric pumping) reduces concurrence.

\begin{figure}[h!]
        \begin{center}
                \noindent
                \includegraphics[width=5.5in]{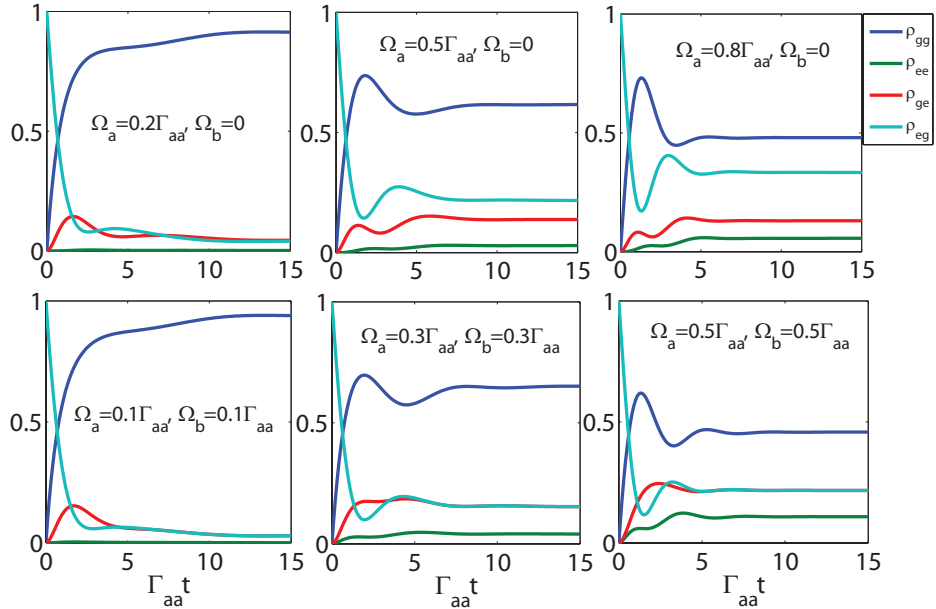}
                \caption{Dynamics of the density matrix elements for qubits under external pumping. Qubits are above a finite groove with qubit separation $ 1.5\lambda_{\mathrm{spp}}=634.5 $ (see Fig. \ref{Fig5} for details of the geometry). Two different regimes of pumping have been considered, symmetric pumping (two field pumping) and asymmetric pumping (single field pumping).}\label{Fig9}
        \end{center}
\end{figure}

Population dynamics of the qubits are shown in Fig. \ref{Fig9} for the case when qubits are separated by $1.5\lambda_{\mathrm{spp}}$ and pumped either symmetrically, or only one qubit was pumped, for different intensities. Four density matrix elements are $\rho_{gg}$, $\rho_{ee}$, $\rho_{ge}$, and $\rho_{eg}$ which are respectively the probability of both qubits to be in ground state, both qubits to be in excited state, the first qubit to be in the ground state and the second to be in the excited state, vice versa. Moreover we have assumed that initial state is a state when $\rho_{eg} = 1$.

The qubits dynamics is different for one-field and two-field pumping. For the case of two-field pumping, in the  steady state we have $ \rho_{eg} = \rho_{ge} $ which is reasonable as we are pumping both systems identically. We can see that these elements first grow rather fast with the growth of the pump intensity, but then slow down around $\Omega_{\mathrm{aa}}=\Omega_{\mathrm{bb}}=0.3\Gamma_{\mathrm{aa}}$. We can also see that the element $\rho_{ee}$ also grows under the influence of the pump. But it first grows slower and then starts growing faster around $\Omega_{\mathrm{aa}}=\Omega_{\mathrm{bb}}=0.3\Gamma_{\mathrm{aa}}$. Therefore, this indicates that under such strong pumping the dynamics of both emitters is mostly defined by external pumps, and not by the qubit-qubit interactions, which is detrimental for entanglement. Under the single field illumination the density matrix elements evolve similarly to the previous case. However, $\rho_{eg} > \rho_{ge}$, because the first qubit that is pumped in this case. We can see that with the intensity increase all elements grow, however, when the intensity becomes too strong the growth freezes and only $\rho_{eg}$ grows, indicating that dynamics of first emitter is entirely defined by the pump field, which is also detrimental for entanglement.

\section{Conclusions}
Transient and steady state entanglement of two qubits mediated by two different types of plasmonic waveguides have been presented. The dynamics of the qubits has been studied using the formalism of the master equations. The waveguides, acting as photonic reservoirs, were incorporated into the formalism through the classical dyadic Green function, accounting for dissipative and coherent interactions between qubits and the reservoir. The Green dyadic has been calculated numerically. Two different methods have been proposed to improve the entanglement between qubits communicating through excited surface plasmonic waves; using a finite, resonant-length plasmonic waveguide, and including coupling slots near the sites of the qubits to improve the coupling of the qubits to the plasmonic modes. The calculated results are compared with the infinite-length versions of the waveguides, and the vacuum case, and it is shown that a finite structure or slotted structure can improve the entanglement of appropriately-positioned qubits.

\section*{Acknowledgements}
This work was supported by the Natural Sciences and
Engineering Research Council of Canada and Queen's University.

\end{document}